\documentclass[10pt,letterpaper]{article}
\usepackage{cogsci_template}
\usepackage{cogsci}
\cogscifinalcopy
\hypersetup{
    citecolor=gray
}
\acrodef{dual}[SHM]{Simulation-Heuristics Model}
\acrodef{ipe}[IPE]{Intuitive Physics Engine}
\acrodef{isr}[ISR]{Integration of Simulation and Rules}

\title{A simulation-heuristics dual-process model for intuitive physics}
\author{
    \begin{tabular}{c c c c}
        \bf Shiqian Li$^{\star,1,2}$\quad{}\quad{} & \bf Yuxi Ma$^{\star,1}$\quad{}\quad{} & \bf Jiajun Yan$^{\star,1}$\quad{}\quad{} & \bf Bo Dai$^{2}$\quad{}\quad{}\\
        \normalfont shiqianli@stu.pku.edu.cn\quad{}\quad{} & \normalfont yxma@stu.pku.edu.cn\quad{}\quad{} & \normalfont yjejuy@gmail.com\quad{}\quad{} & \normalfont daibo@bigai.ai\quad{}\quad{}
    \end{tabular}\vspace{2pt}
    \\
    \begin{tabular}{c c c}
        \bf Yujia Peng$^{1,2,3,\,\textrm{\Letter}}$\quad{}\quad{} & \bf Chi Zhang$^{1,\,\textrm{\Letter}}$\quad{}\quad{} & \bf Yixin Zhu$^{1,\,\textrm{\Letter}}$\quad{}\quad{}\\
        \normalfont yujia\_peng@pku.edu.cn\quad{}\quad{} & \normalfont wellyzhangc@gmail.com\quad{}\quad{} & \normalfont yixin.zhu@pku.edu.cn\quad{}\quad{}
    \end{tabular}\vspace{2pt}
    \\
    \begin{tabular}{r l}
        \small $\star$ equal contributors\quad{} & $\textrm{\Letter}$ corresponding authors\\
        \small $^1$ Institute for Artificial Intelligence, Peking University&
        \small $^2$ State Key Laboratory of General Artificial Intelligence, BIGAI
    \end{tabular}
    \\
    \small $^3$ School of Psychological and Cognitive Sciences and Beijing Key Laboratory of Behavior and Mental Health, Peking University \\ \vspace{3pt}
    \url{https://github.com/lishiqianhugh/DualModel} \\ \vspace{2pt}
}
\begin{document}
\maketitle

\begin{abstract}
The role of mental simulation in human physical reasoning is widely acknowledged, but whether it is employed across scenarios with varying simulation costs and where its boundary lies remains unclear. Using a pouring-marble task, our human study revealed two distinct error patterns when predicting pouring angles, differentiated by simulation time. While mental simulation accurately captured human judgments in simpler scenarios, a linear heuristic model better matched human predictions when simulation time exceeded a certain boundary. Motivated by these observations, we propose a dual-process framework, \ac{dual}, where intuitive physics employs simulation for short-time simulation but switches to heuristics when simulation becomes costly. By integrating computational methods previously viewed as separate into a unified model, \ac{dual} quantitatively captures their switching mechanism. The \ac{dual} aligns more precisely with human behavior and demonstrates consistent predictive performance across diverse scenarios, advancing our understanding of the adaptive nature of intuitive physical reasoning. 

\end{abstract}

\begin{figure*}[t!]
    \centering
    \includegraphics[width=\linewidth]{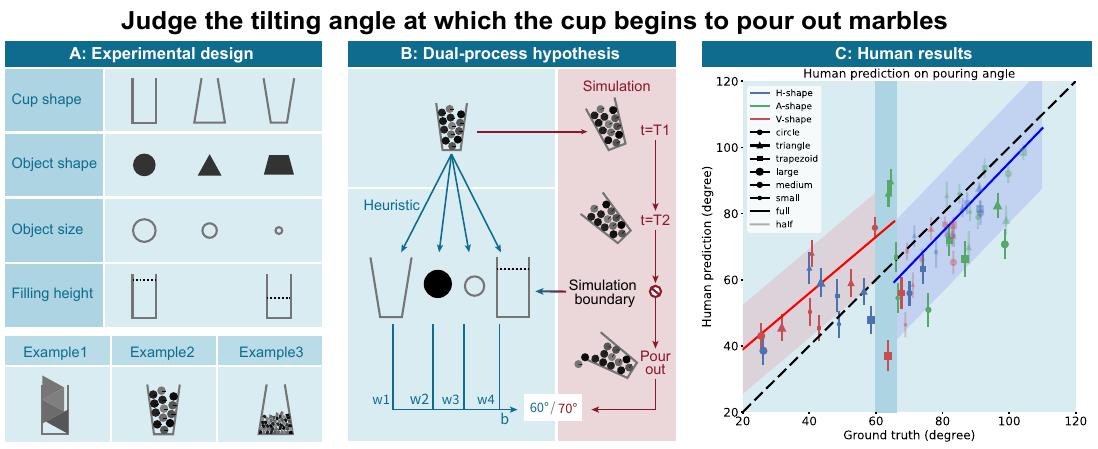}
     \caption{\textbf{(A) Experimental design:} Trials involved 3 cup shapes (H-shape, A-shape, V-shape), 3 object shapes (circle, triangle, trapezoid), 3 sizes (large, medium, small), and 2 filling heights (full, half), totaling 54 unique conditions. Participants predicted the tilt angle for marbles to fall out when cups are tilted to the left. \textbf{(B) \ac{dual} hypothesis:} Participants used either mental simulation, simulating the tilting process until pouring out, or a heuristic strategy, reaching judgments from physical features when the simulation exceeds a boundary. These methods could result in different outcomes. \textbf{(C) Human results:} Each point represents a condition, illustrating human tendencies to either overestimate or underestimate the pouring angle. The red and blue lines are the regression results of \ac{ipe} and the heuristic model, respectively. The \ac{dual} effectively captures human behavior with a switching boundary.}
    \label{fig:overview}
\end{figure*}

\section{Introduction}\label{sec:intro}

Humans demonstrate extraordinary abilities in understanding and reasoning about the physical world even without formal training in physics \citep{piloto2022intuitive}. This ability, known as intuitive physics \citep{kubricht2017intuitive}, enables comprehending physical concepts \citep{baillargeon1985object,baillargeon1987s,kim1992infants}, predicting physical dynamics \citep{battaglia2013simulation,bates2015humans,davis2017commonsense}, and interacting with the physical environments \citep{allen2020rapid}. However, human intuitive physics may exhibit errors and biases in certain physical scenarios, indicating deviations from classical Newtonian physics \citep{mccloskey1980curvilinear,mccloskey1983intuitive,kaiser1986intuitive,kozhevnikov2001impetus}. Such errors and biases serve as a unique aspect of human reasoning, offering a valuable avenue for studying the underlying mechanisms of intuitive physics \citep{kubricht2017intuitive}.

A common perspective to understanding human intuitive physics is mental simulation: it hypothesizes an approximate intuitive physics engine in the human mind \citep{battaglia2013simulation,smith2013sources,ullman2017mind,smith2024intuitive}. This simulation framework, grounded in probabilistic inference, was found to be able to characterize human behavior across various physical tasks, and also account for human errors and biases, further validating its relevance and applicability \citep{battaglia2012computational,kubricht2016probabilistic,kubricht2017consistent,gerstenberg2017faulty,ullman2018learning,bates2019modeling,bass2021partial,chen2023just,li2023approximate}. Nevertheless, the simulation model fails to completely explain the variance in human behavior in some demanding or unfamiliar conditions \citep{schwartz1999inferences,kozhevnikov2001impetus,smith2018different,ludwin2020broken,ludwin2021limits}, suggesting the existence of alternative cognitive mechanisms, possibly mental shortcuts employed for certain physical scenarios, or heuristics \citep{davis1998naive,davis2017commonsense,kozhevnikov2001impetus,kubricht2017intuitive,smith2018different}. Recent work has further explored these limitations, showing that mental simulation has severe capacity constraints. \citet{balaban2024capacity} found that people can track only a single moving object in imagination. \citet{li2023approximate} showed that people use simplified object approximations for physical reasoning, employing shortcuts that reduce computational demands. Particularly relevant to our work, \citet{ullman2023resource} identified a switch point in a fluid-reasoning task where prediction patterns changed based on simulation demands. These findings in capacity constraints raise important questions about how humans might adapt their reasoning strategies when faced with increasingly complex physical scenarios.

Here we ask the questions: \textit{Do humans consistently rely on mental simulation, or do they employ alternative heuristic strategies under certain conditions? What are the situations that prompt a switch between these two cognitive strategies?}

Previous studies have investigated the interplay between simulation and heuristics, providing evidence for \textbf{qualitative} insights. For instance, \citet{kozhevnikov2001impetus} demonstrates that people tend to use impetus heuristics in quick judgment scenarios, while \citet{battaglia2013simulation} finds that models based on height heuristics can more accurately explain human judgment in certain tasks, such as predicting the falling distance of a block tower. Furthermore, \citet{smith2017thinking} and \citet{yildirim2017physical} suggest that humans integrate these two cognitive strategies in motion prediction and physical planning tasks, respectively. However, there is currently no study that has provided clear evidence supporting the relationship between these two strategies or \textbf{quantitatively} demonstrated the transition between them. A comprehensive exploration is needed to understand whether a switch of policies exists and, if so, how these switches operate, as well as to identify alternative heuristics that could reverse engineer the human physical reasoning process, including human biases.

In our study, we systematically investigate the switch between simulation and heuristic strategies in intuitive physics, developing a computational model that offers improved explanatory power. We hypothesize that: (i) the simulation strategy prevails in scenarios simple enough for reliable physical unfolding; (ii) the heuristic strategy takes over when mental simulation becomes too costly; (iii) the switching point of the two strategies correlates with the simulation cost, approximated via a proxy of simulation time. Diverging from previous studies that often focused on simpler dynamics or predictable outcomes \citep{battaglia2013simulation,smith2013sources,smith2017thinking}, our study engages in examining human reasoning across a range of simulation costs \citep{schwartz1999inferences,kubricht2016probabilistic,davis2017commonsense}. Inspired by previous pouring tasks in intuitive physics \citep{schwartz1999inferences,kubricht2016probabilistic,guevara2017adaptable,lopez2020stir}, we build a pouring marble task with more diverse physical properties and complexities. In this task, human participants are asked to judge the tilt angle needed to pour marbles from cups under various setups (see \cref{fig:overview}A). 

We conducted four steps of experiment to validate the above hypotheses sequentially. The \textbf{first step} examines whether there is a pattern switch regarding human judgment. A finding of two distinct error patterns (\ie overestimation and underestimation) supports the existence of two predominant strategies that vary under different simulation times. The \textbf{second step} aims to test our hypothesis on whether the \ac{ipe} model can account for human judgments in simpler scenarios. The results show that it aligns well with human judgments and exhibits the same overestimation when the actual pouring angle is small. However, it fails to account for humans' underestimation as the actual pouring angle exceeds a certain boundary (see \cref{fig:overview}C). Given that the pouring rate remains consistent, we hypothesize longer simulation time leads to increased cost of physical unfolding, triggering the transition to another cognitive strategy. Thus, we validate our second hypothesis by exploring an alternative heuristic approach in the \textbf{third step}. We developed a linear heuristic model trained on ground-truth data and found that, although less effective than \ac{ipe} at smaller angles, the model accurately captures the underestimation pattern when the pouring angle exceeds a certain boundary. These results support our hypothesis of a cognitive shift to a heuristic strategy. To test our third hypothesis, in the \textbf{fourth step}, we explore whether a novel framework, \acf{dual}, that combines these two models and toggles based on simulation cost, can explain human judgments across all complexity levels (see \cref{fig:overview}B). The results show that \ac{dual} aligns more closely with human behavior across diverse scenarios and metrics, enhancing our understanding of intuitive physical reasoning and highlighting the adaptability and versatility of human cognition.

\section{Models}\label{sec:model}

\subsection{Mental simulation}\label{sec:ipe}

Recent work explains human intuitive physics understanding by assuming an approximate simulation engine in the human mind \citep{battaglia2013simulation,lake2017building,kubricht2016probabilistic}. This engine serves to simulate the future physical unfolding, akin to a computational physics engine but incorporates noise into the physical properties of objects.

Following this approach, our model utilizes an \ac{ipe} that runs noisy simulations as in \citet{battaglia2013simulation}. The model takes an initial physical scene $S_{0}$ and external forces $f_{0:T-1}$ to derive the judgment $J$. This process involves predicting the intermediate states $S_{1:T}$ over a time span $T$:
\begin{equation}
    P(J | S_{0}, f_{0:T-1}) =  \int_{S_{1:T}} P(J | S_{1:T}) P(S_{1:T} | S_{0}, f_{0:T-1}) dS_{1:T},
\end{equation}
where $S_{1:T}$ represents the sequence of all physical states from time steps 1 through T, and the integral averages over all possible trajectories of these states.
Each state evolves as $S_{t+1} = \phi(S_{t} + \epsilon, f_{t})$ with noise $\epsilon \sim \mathcal{N}(0, \sigma^2)$, and $\phi(\cdot)$ representing deterministic physical dynamics. We simplify the mapping from the initial state to the final judgment as $M(S_{0}; f, \epsilon)$.

In our implementation using the flexible physics engine Pymunk, the \ac{ipe} utilizes all physical variables to simulate future dynamics in a 2D scenario with added Gaussian noise $\mathcal{N}(0, \sigma^2)$ to each marble's position horizontally and vertically during every simulation step. When this noise causes objects to overlap, Pymunk's collision resolution automatically repels them based on physical constraints. The noise level $\sigma^2$ is varied from 0.1 to 1 to observe its impact on the simulation results. Inspired by classical findings by \citet{shepard1971mental}, who demonstrated that mental rotation time is proportional to rotation angle, suggesting a constant angular velocity in mental simulation. We assume a simulation model with constant rotational speed and optimize it through iterative parameter search to minimize the RMSE between model predictions and human judgments across a calibration subset of trials. After optimization, the simulation time becomes proportional to the rotation angle.  \textit{In the following context, we will use rotation angles to represent simulation time}.

We perform 30 noisy \ac{ipe} simulations per trial. During each simulation, an automatic detection system is integrated to identify the moment when the marbles fall out, which serves as the ground truth. The final pouring angle is determined from the average of the 30 results. This setup allows us to mimic the variability and uncertainty in human cognition, as outlined in prior studies \citep{smith2013sources}, and to explore how these factors influence judgment in physical tasks.

\subsection{Heuristic model}

Prior studies often employ predefined heuristics to elucidate human biases \citep{schwartz1999inferences,kozhevnikov2001impetus,smith2017thinking} or fit heuristic models on human data to evaluate the influence of physical attributes \citep{zhou2023mental}. While these approaches offer insights for specific tasks, a systematic methodology for learning heuristics in complex scenarios is lacking.

Our heuristic model is designed to learn from a subset of physical attributes, fitting ground-truth data through a direct mapping $g$ from the initial scene $S_{0}$ to the final judgment $J$, bypassing the intermediate states. This model is advantageous as it approximates humans' real-world physics understanding by a limited set of attributes, and circumvents the need for computationally heavy physics simulation. In particular, we employ a linear model with learnable parameters:
\begin{equation}
    J = g(S_{0}^{1},... , S_{0}^{n}) = \sum_{i=1}^{n} \omega_{i}S_{0}^{i} + b,
\end{equation}
where $\{S_{0}^{i}\}$ are different physical variables in $S_{0}$ and we set $n = 4$ in our study. Specifically, the model considers the following four variables: object size, filling height, object shape, and cup shape. Instead of directly predicting the pouring angle, the model predicts the difference between the actual pouring angle and a reference 90-degree angle. This design choice was made based on preliminary observation from a familiarization experiment that an H-shape cup containing little marbles almost always pours out at 90 degrees. The model is optimized using the mean squared error. Future exploration may consider nonlinear heuristic models using symbolic regression \citep{xu2021bayesian}.

\subsection{Dual-process model}

Building on the notion that human cognition might employ multiple systems \citep{kahneman2011thinking}, we introduce a dual-process model in the context of intuitive physics, termed \ac{dual}. This model hypothesizes that humans alternate between two strategies, mental simulation and heuristic reasoning, based on the simulation cost, which is approximated by the simulation time. The simulation strategy parallels Kahneman's "slow thinking" (System 2): a deliberate, resource-intensive process, while the heuristic approach functions as "fast thinking" (System 1): an efficient but potentially less accurate mechanism. Specifically, for simulation time below a critical boundary $\theta$, \ac{ipe} is favored, whereas, beyond $\theta$, a heuristic strategy is triggered. This adaptive approach is formalized as:
\begin{equation}
    \begin{aligned}
        \begin{cases}
            J = E_\epsilon[M(S_{0}; \epsilon)], & \text{if } T \leq \theta,\\
            J = \sum_{i=1}^{n} \omega_{i}S_{0}^{i} + b, & \text{if } T > \theta
        \end{cases}.
    \end{aligned}
\end{equation}
where we drop the dependency on $f$, which remains constant across the same set of experiments. We employ a grid search method to optimize both $\theta$ for the strategic transition and the noise parameters $\sigma$ for the \ac{ipe}, in addition to a group of heuristic parameters $\omega$ derived from linear regression.

\section{Experiment}

\subsection{Participants}\label{sec:participant}

A total of 43 college students (55\% male, 45\% female; mean age = 21.77 ± 4.45) completed in-person experiments for course credit or monetary compensation. One participant was excluded due to little response variation. The study received IRB approval from the Committee for Ethics and the Protection of Human and Animal Welfare at the local institution.

\subsection{Stimuli}

Stimuli were generated using Pymunk in various configurations, encompassing three cup shapes (H-shape, A-shape, and V-shape), three object shapes (circle, triangle, and trapezoid), three object sizes (large, medium, and small), and two filling heights (half and full), totaling up to 54 different conditions. These stimuli were then rendered using Pygame.

Marbles were randomly placed inside cups, with layouts adjusted by Pymunk's physics engine. Three random layouts were generated per condition. Stimuli were selected when marbles reached stability. Marbles had no friction or elasticity, equal mass, and random grayscale colors to eliminate material property assumptions.

\begin{figure}[ht]
    \centering
    \begin{subfigure}{0.57\linewidth}
        \centering
        \includegraphics[width=\linewidth]{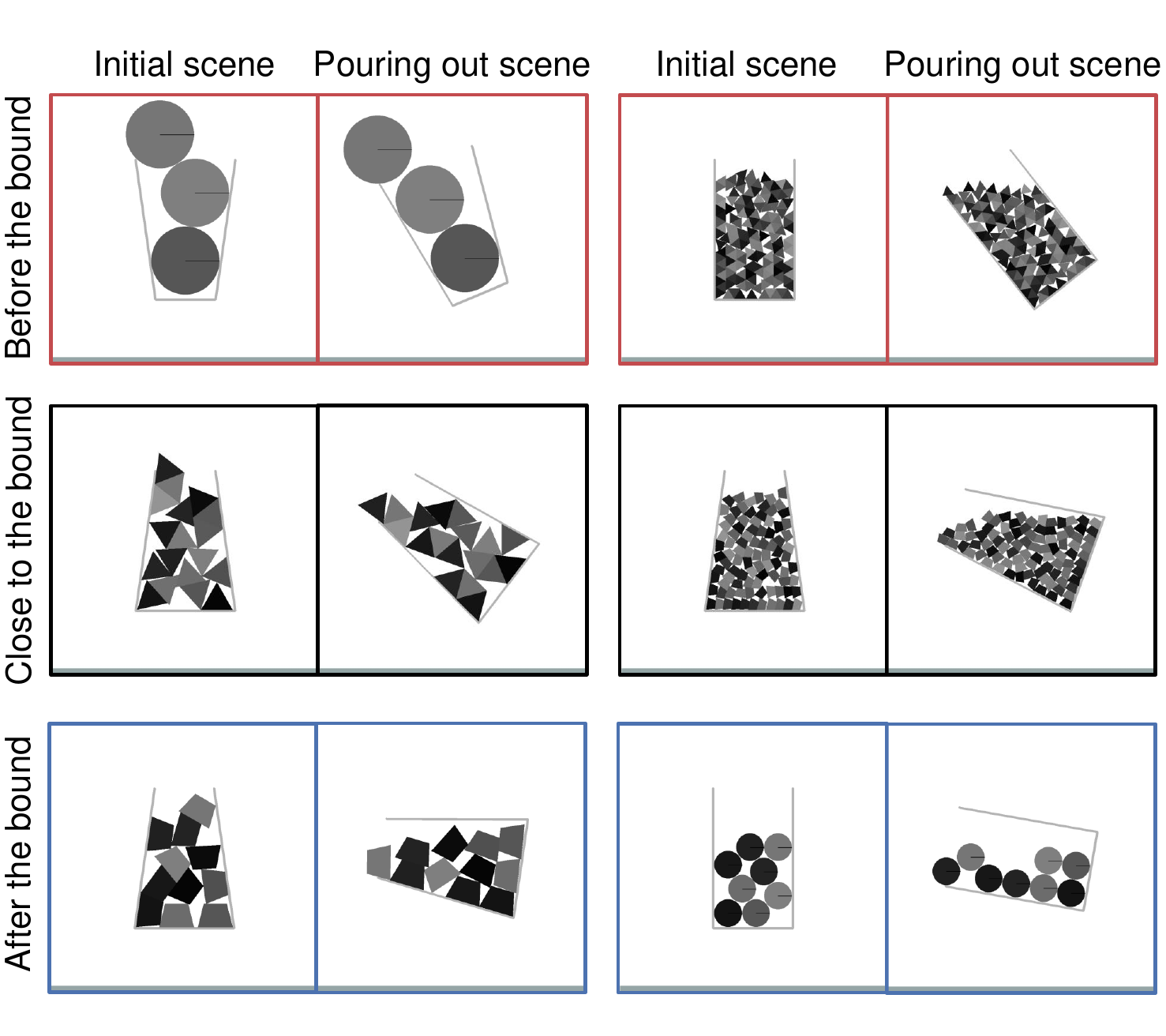}
        \caption{}
        \label{fig:stimuli}
    \end{subfigure}%
    \begin{subfigure}{0.43\linewidth}
        \centering
        \includegraphics[width=\linewidth]{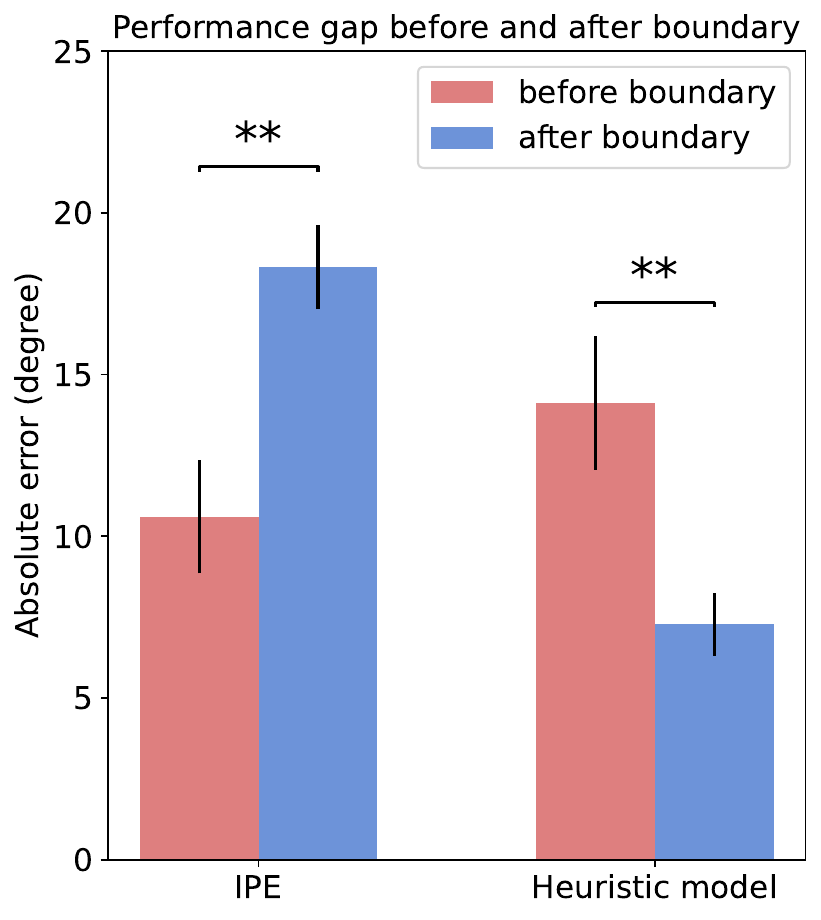}
        \caption{}
        \label{fig:abs_err}
    \end{subfigure}
    \caption{Visualizations of stimuli and error analysis. (a) Example stimuli. The top (red), middle (black), and bottom (blue) rows depict two scenarios each, with pouring angles that are smaller, close to, and larger than the established simulation bound, respectively. (b) The mean absolute error between model and human results (with SEM). The \ac{ipe} model exhibits a larger absolute error when the simulation time exceeds the boundary. Conversely, the heuristic model shows contrary results, indicating its effectiveness in these scenarios.}
    \label{fig:combined}
\end{figure}

Pouring angles were determined through simulations of slow cup rotations. The critical angle was measured when a marble's mass center aligned with the cup's top-left corner, using 120 FPS calculations with automatic fall detection. Tilt angles were referenced to verify pouring angles. Example stimuli appear in \cref{fig:stimuli}.

\subsection{Procedures}

A within-subjects design was implemented, where each participant completed all 54 conditions. Stimuli were navigated in a counter-balanced order with randomly selected layouts, and the experiment lasted approximately 30 minutes.

\paragraph{Familiarization}

Following informed consent, participants underwent an instruction phase and familiarization session featuring pouring demonstrations with two small marbles, addressing marble properties, tilt angles, and pour initiation criteria. Comprehension was assessed through quizzes on empty cup angles and marble pour timing before proceeding to the experimental phase.

\paragraph{Experiment}

Participants were required to complete 54 trials consecutively. In each trial, a static image of a non-rotated cup from various setups was presented. The tilt angle necessary for the cup to begin pouring out marbles was estimated by the participants using a slider bar with a range of 0 to 135 degrees. To reduce potential biases from inaccurate angle perception, a dial marked with angle measurements was provided in each trial. Demographic information along with the responses for the pouring angles across all 54 trials, including the total duration, were recorded for subsequent analysis.

\section{Results}\label{sec:result}

Our analysis follows four sequential steps to validate our hypothesis: (1) examination of participants' error patterns indicating strategic shifts, (2) evaluation of \ac{ipe}'s capacity to explain human judgment, (3) development of a complementary heuristic model incorporating physical attributes, and (4) integration of simulation and heuristic approaches into a hybrid model to comprehensively explain human judgments.

\subsection{A switching in error patterns}
Human results show overestimation and underestimation of the pouring angle compared with the ground truth. These two error patterns may indicate different strategies of physical reasoning. To examine whether there is a switching mechanism between the two patterns among those conditions, we employed symbolic regression to automatically identify an explainable factor and its corresponding switching point that best distinguishes between the two patterns. We considered all experimental design factors, including cup shape, object shape, object size, and filling height, along with the object number and simulated pouring angle. Our analysis shows that the simulated pouring angle effectively differentiates between the reversal patterns observed in human participants' estimations of tilt angles for pouring (see \cref{fig:overview}C). We identified the optimal boundary for distinguishing these patterns to be 65 degrees by searching from 20 to 120 with an interval of 1. Initially, participants tended to overestimate these angles when the simulated pouring angles were relatively small (mean discrepancy = 7.76 ± 13.67). As the angle increases, this trend shifts to consistent underestimation (mean discrepancy = -9.89 ± 8.75). Given the consistent tilting speed, the observed pattern switch as the pouring angle increases suggests a hypothesis that the physical reasoning strategy may change when the simulation time exceeds a certain resource boundary.

\subsection{\texorpdfstring{\ac{ipe}}{} fails to explain all trials}

 To validate our hypothesis, we first experiment with the \ac{ipe} model. Fitting human judgments in the overestimation phase with \ac{ipe} supports our hypothesis of the simulation strategy's dominance in the shorter time span. Note that as the angular speed remains constant in our experiments, the simulation time is proportional to the degree of angle. When the positional noise and rotational speeds of the \ac{ipe} model were optimized, the results were closely aligned with human performance, explaining the overestimation pattern effectively (r = .890).

However, once the pouring angle exceeded the 65-degree boundary, \ac{ipe}'s prediction error significantly increased (t(52) = -3.354, p = .002; see \cref{fig:abs_err} for absolute error comparison on the left). No parameter combination in the \ac{ipe} model could well explain the underestimation pattern, indicating the existence of an alternative strategy other than \ac{ipe}.

\subsection{Learned heuristic model complements \ac{ipe}}

\begin{table}[ht!]
    \centering
    \small
    \caption{\textbf{Categories, coefficients, and p-values of physical variables in the learned heuristic model.} All physical variables except the object shape show significant contributions to the outcomes.}
    \label{tab:heu}
    \begin{tabular}{cccc}
        \toprule
        \textbf{Variable} & \textbf{Category} & \textbf{Coefficients} & \textbf{p} \\ 
        \midrule
        \multirow{3}{*}[-1ex]{Cup shape} & H-shape & \multirow{3}{*}[-1ex]{-11.528} & \multirow{3}{*}[-1ex]{0.000} \\ 
        \cmidrule(lr){2-2}
        & A-shape & & \\ 
        \cmidrule(lr){2-2}
        & V-shape & & \\ 
        \midrule
        \multirow{3}{*}[-1ex]{Object shape} & Circle & \multirow{3}{*}[-1ex]{1.577} & \multirow{3}{*}[-1ex]{0.073} \\ 
        \cmidrule(lr){2-2}
        & Triangle & & \\ 
        \cmidrule(lr){2-2}
        & Trapezoid & & \\ 
        \midrule
        \multirow{3}{*}[-1ex]{Object size} & Small & \multirow{3}{*}[-1ex]{7.029} & \multirow{3}{*}[-1ex]{0.000} \\ 
        \cmidrule(lr){2-2}
        & Medium & & \\ 
        \cmidrule(lr){2-2}
        & Large & & \\ 
        \midrule
        \multirow{2}{*}[-0.5ex]{Filling height} & Half & \multirow{2}{*}[-0.5ex]{-19.955} & \multirow{2}{*}[-0.5ex]{0.000} \\ 
        \cmidrule(lr){2-2}
        & Full & & \\ 
        \bottomrule
    \end{tabular}
\end{table}

To better explain the underestimation pattern in human behavior, we devised a heuristic model incorporating key physical attributes rooted in our experiments: filling height, cup shape, object shape, and object size. This model effectively compensated for the discrepancies unexplained by the \ac{ipe} model. The heuristic model performed well when the actual pouring angle exceeded 65 degrees (r = .841), but its accuracy diminished below this boundary (Mann-Whitney U test, p = .003; see \cref{fig:abs_err} for absolute error comparison on the right).

Further analysis of specific heuristics revealed that filling height, cup shape, and object size significantly influence heuristic judgment (see \cref{tab:heu}, p = .000 for all three variables). The model's coefficients allowed a quantitative assessment of these variables' impact.  Despite the simplicity and approximate encoding, this linear heuristic model captured basic physical intuition effectively. The findings align with our second hypothesis, suggesting the adoption of heuristic strategies when mental simulation reaches its boundary.

\subsection{\texorpdfstring{\ac{dual}}{} explains human judgments on all conditions}

Building upon our findings, we constructed the \acf{dual}, a dual-process model integrating both simulation and heuristic strategies, to optimally predict human performance across all trials. Instead of relying on actual simulation time in humans, which is unavailable, we instead based the transition criterion in \ac{dual} on \ac{ipe}'s simulation time. A grid search identified the boundary of 68.2 degrees in simulation time and a dynamic positional noise of 0.2 as optimal for mirroring human judgments.

\begin{figure}[t!]
    \centering
    \includegraphics[width=\linewidth]{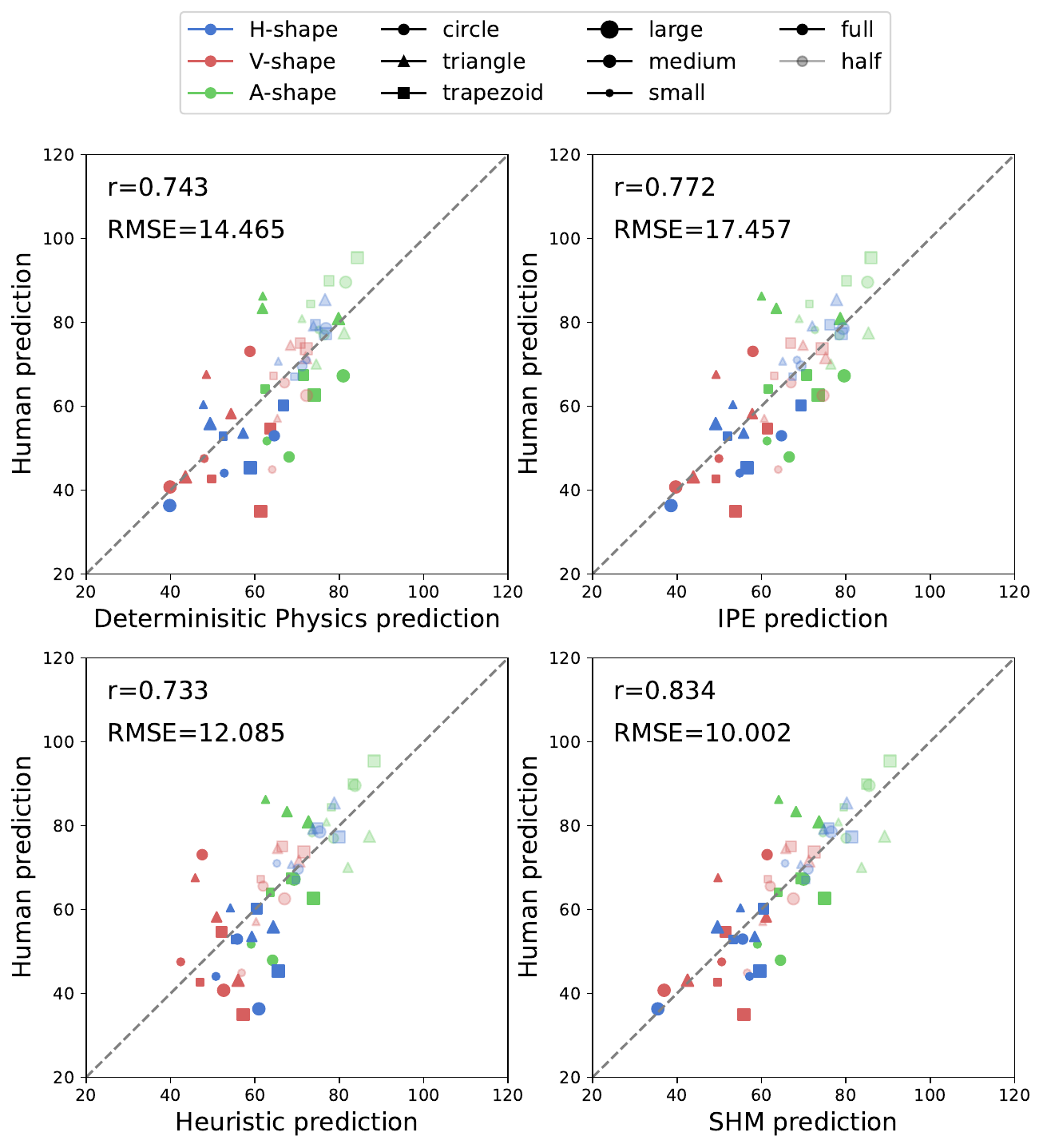}
    \caption{\textbf{Comparison between \ac{dual} and other baseline models.} The correlation and RMSE between model predictions and human predictions across all 54 conditions are compared. Among the four models evaluated, \ac{dual} demonstrates the highest correlation and the lowest RMSE, indicating its superior predictive accuracy.}
    \label{fig:r}
\end{figure}

In predicting overall human performance, \ac{dual} surpassed three baseline models: the deterministic physics model, \ac{ipe}, and the purely heuristic model. \ac{dual} exhibited the highest correlation and lowest RMSE (r = .834, RMSE = 10.002), as shown in \cref{fig:r}. Although \ac{ipe} was correlated with human judgments (r = .772), it showed high error in making human-like predictions (RMSE = 17.457). On the contrary, the heuristic model could predict human judgments with smaller RMSE but failed to better explain the variance (r = .733, RMSE = 12.085).

\begin{figure*}[t!]
    \centering
    \includegraphics[width=\linewidth]{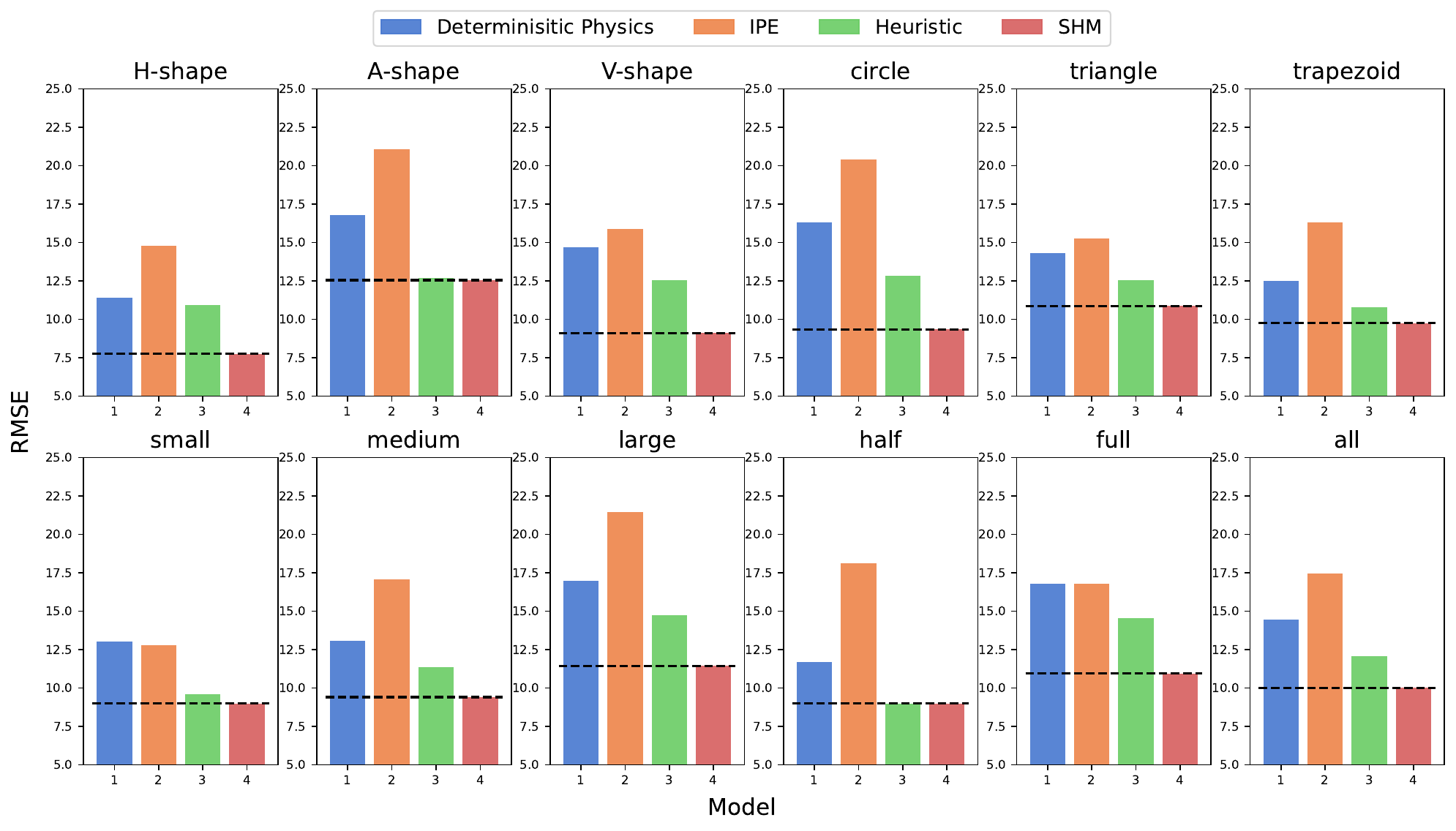}
    \caption{\textbf{Comparison of four models' RMSE on different conditions.} RMSE is calculated as the root mean square error between the model's predicted pouring angle and the human judgments. The bottom right figure represents the performance across all 54 trials. A dashed line is included to indicate the RMSE of the SHM, showing a clear advantage when compared with other models.}
    \label{fig:rmse}
\end{figure*}

The fitted \ac{dual} model exhibited consistent predictive performance across diverse conditions (\eg, different cup shapes, object shapes, sizes, and filling heights). It consistently showed the lowest RMSE, except in specific scenarios where the heuristic model was parallel (\cref{fig:rmse}). The model explained maximum variance in almost all cases, with comparable performance to \ac{ipe} in scenarios involving large or trapezoidal marbles. Notably, in situations where \ac{ipe} minimally correlated with human judgments (\eg, A-shaped cups, r = .461), \ac{dual} maintained effectiveness (A-shaped cups, r = .647). It also significantly improved correlation in scenarios poorly addressed by the heuristic model (full filling height, r improved to .673 from .377). These results highlight \ac{dual}'s capability to synergize the strengths of both \ac{ipe} and the heuristic model, enabling robust predictions across diverse scenarios. Consequently, the \ac{dual} model, with its transition mechanism based on simulation time, aligns with our third hypothesis and effectively accounts for a wide range of conditions and metrics.

\section{Discussion}

\paragraph{Limitations}\label{sec:limit}
While our study proposes a method to distinguish between cognitive strategies, people can employ multiple strategies within a single trial. Furthermore, existing research \citep{siegler1976three,marcus2013robust,davis2016scope,ludwin2021limits} in intuitive physics indicates that humans' results can be incompatible with those from mental simulation models, even for simple tasks like predicting the tipping direction of the balanced beams. This suggests that factors other than the mental cost triggered the heuristic strategy, and a more nuanced modeling is needed to capture the variability in human judgment.
We note that our model parameters were fitted using the complete dataset rather than through cross-validation on separate subsets of conditions. Therefore, our claims focus on the model's ability to explain human behavior consistently across diverse physical scenarios rather than its ability to predict novel conditions.

\paragraph{Future work}
Future research could investigate the mechanisms behind strategy selection and the generalizability of our findings across various physical scenarios. Key questions include how humans choose an initial strategy and estimate simulation costs. Eye-tracking or verbal protocol studies might reveal whether people start with simulation and abandon it when costly, or make preemptive decisions based on scene complexity. Although our use of pouring angles is task-specific, the underlying cost-based strategy-switching mechanism likely applies more broadly.

Additionally, the dual-process model employed by humans may utilize different physical variables as heuristics across scenarios. For instance, tower height might serve as a heuristic in block collapse estimation, while object distribution could be relevant for predicting group motion. Other factors influencing strategy selection might include task complexity, scenario familiarity, and prior experience. Despite this variance in heuristic strategies, investigating whether our learning approach maintains effectiveness across contexts would be valuable for developing a deeper understanding of physical reasoning.

\section{Conclusion}

In this work, we design a pouring-marble task to study the computational mechanism in intuitive physics. The sequential experiments underscore that while the \ac{ipe} effectively predicts human judgments in scenarios with short simulation times, its efficacy diminishes as these times extend. This limitation of \ac{ipe} paves the way for the implementation of a heuristic approach that shows greater accuracy in scenarios necessitating longer and more complex simulations. The introduction of the \ac{dual} model, which seamlessly integrates these two cognitive strategies based on the simulation cost, not only aligns more closely with human behavior but also enhances the model’s consistent predictive capabilities across varied conditions. By bridging the gap between mental simulations and heuristic approaches, the \ac{dual} model offers a robust framework to capture the complexity and adaptability of human cognition in intuitive physics. This model serves as a pivotal step in exploring computational methods that closely mimic human-like reasoning, providing insights into the cognitive mechanisms that govern our interactions with the physical world.

\paragraph{Acknowledgement}

The authors would like to thank Miss Chen Zhen (BIGAI) for making the nice figures, and Prof. Yizhou Wang (Peking University) for the helpful discussion. This work is supported in part by the National Natural Science Foundation of China (62376031), the State Key Lab of General AI at Peking University, the PKU-BingJi Joint Laboratory for Artificial Intelligence, and the National Comprehensive Experimental Base for Governance of Intelligent Society, Wuhan East Lake High-Tech Development Zone.

\bibliographystyle{apacite}
\renewcommand\bibliographytypesize{\small}
\setlength{\bibleftmargin}{.125in}
\setlength{\bibindent}{-\bibleftmargin}
\bibliography{reference_header,reference}

\end{document}